\documentclass[aps,prl,onecolumn,superscriptaddress,floatfix,preprint]{revtex4}
\usepackage{epsfig,amsmath,amssymb,color}
\begin{document}
\title{The Mott metal-insulator transition in the 1D Hubbard model in an external
  magnetic field}
 
\author{Holger Frahm} 
\affiliation{Institut f\"ur Theoretische Physik, Leibniz Universit\"at
  Hannover, Appelstra\ss{}e 2, 30167 Hannover, Germany}
\author{Temo Vekua}
\affiliation{Laboratoire de Physique Th\'eorique et Mod\`eles
Statistiques, Universit\'e Paris Sud, 91405 Orsay Cedex, France}

\date{04. December 2007}
    
\begin{abstract}
We study the low energy behavior of the one dimensional Hubbard model across
the Mott metal-insulator phase transition in an external magnetic field. In
particular we calculate elements of the dressed charge matrix at the critical
point of the Mott transition for arbitrary Hubbard repulsion and magnetization
numerically and, in certain limiting cases, analytically.  These results are
combined with a non-perturbative effective field theory approach to reveal how
the breaking of time reversal symmetry influences the Mott transition.
\end{abstract}
\maketitle

\section{introduction}
Exact solutions play an invaluable role in our understanding of the behavior
of electron systems in one dimension, especially for effective field theory
approaches in regimes when strong coupling develops. Here effective theories
have to be assisted with a non-perturbative, unbiased analysis. Strong
coupling regimes can develop when the initial interactions between electrons
are strong, but also in certain situations when the bare couplings are
arbitrarily weak. One well known example of the latter case is provided by the
so called commensurate-incommensurate phase transition \cite{JN,PT}, 
in which a single bosonic mode is involved.  By comparing with a free fermions picture the universal
physics at long wavelengths across the transition was established \cite{JN}.
A physical system realizing such transition is e.g. the repulsive Hubbard
model at half filling, where Mott metal-insulator transition takes place upon
variation of the chemical potential in the absence of an external magnetic
field.  In this case the charge mode undergoes a commensurate-incommensurate
phase transition, while the spin sector is decoupled.

In the absence of special symmetries such as time reversal symmetry, however,
there is no reason for the decoupling of spin and charge modes.  This is well
known for the Hubbard model away of half filling in finite magnetic field
\cite{Frahm}.
At half-filling, on the other hand, spin and charge modes are strictly
decoupled even in an external magnetic field and the low energy sector is
equivalent to the Heisenberg antiferromagnet.  Here one might be led to think
that the admixture between spin and charge modes will die out gradually approaching the half filling.

If this expection would be correct, the Mott metal-insulator transition in the
presence of an external magnetic field would look pretty similar to that at
zero field: all relevant changes affect only the charge sector, while the spin
sector remains practically undisturbed.  Thus the phase transition would fall
in the the single mode commensurate-incommensurate universality class 
even in the presence of magnetic field. In the following we will
show that the above expectation is misleading. Any small coupling between the
spin and charge modes increases under the renormalization and qualitatively influences the critical
properties of the system.

\section{The model}
Our starting microscopic model is the Hubbard Hamiltonian of one dimensional
lattice electrons:
\begin{eqnarray}
H&=&-\sum^N_{i=1,\sigma}(c_{i,\sigma}^{\dagger}c_{i+1,\sigma}+c_{i+1,\sigma}^{\dagger}c_{i,\sigma})+4u\sum^N_{i=1}(\frac{1}{2}-n_{i,\downarrow})(\frac{1}{2}-n_{i,\uparrow})\nonumber\\
&-& \mu \sum^N_{i=1}(n_{i,\uparrow}+n_{i,\downarrow})
-\frac{h}{2}\sum^N_{i=1}(n_{i,\uparrow}-n_{i,\downarrow})
\label{hamil}
\end{eqnarray}
Here $n_{i,\sigma}$ is the number operator of electrons with spin
$\sigma=\{\uparrow, \downarrow\}$ at the $i$-th site, $4u>0$ is the local
coupling constant, $\mu$ is the chemical potential (half filling corresponds
to $\mu=0$), and $h$ is an external magnetic field.

The asymptotics of the correlation functions of Hubbard model in the presence
of two gapless modes have been computed using a conjecture for a multivelocity
'conformal' field theory \cite{IzKR89} by finite size scaling analysis of the
low lying energies of (\ref{hamil}) from the Bethe ansatz
\cite{FrKo90,Frahm}. Alternatively, the exact finite size spectrum can be used
to determine the parameters of a multi-component Luttinger liquid theory in a
non-perturbative way. This way one obtains an explicit expression for the
correct low energy effective field theory for the Hubbard model in the sector
corresponding to the given filling and magnetic field \cite{Penc}.
In both approaches, the crucial quantity governing the low energy behavior is
the $k\times k$ dressed charge matrix, $k$ being the number of gapless modes
in the theory.

We will first try to follow the second approach and work out the effective
field theory to better understand non-perturbative effects produced by the
magnetic field across Mott metal-insulator transition. Then we will compute
elements of dressed charge matrix from the Bethe ansatz, numerically for
generic Hubbard coupling and magnetic fields, and in certain limits
analytically.  Finally we will study response functions and correlation
functions and discuss the effects of the magnetic field on the Mott metal
insulator transition.
\section{effective field theory}
\subsection{Weak-coupling Bosonization}
First we recapitulate on the effective field theory of the repulsive Hubbard
model at half filling where umklapp processes open a charge gap and
the low energy sector of the Hubbard model becomes equivalent to that of the
Heisenberg antiferromagnetic chain. When the chemical potential is smaller
than the gap in the single particle excitation spectrum (charge gap) the
effective Hamiltonian takes the following form \cite{Gogolin}:
\begin{eqnarray}
\label{efHam}
H&=&\frac{v_s}{2}\left[ K_s(\partial_x \theta_s)^2 +(\partial_x\phi_s)^2/K_s \right ] +\frac{v_c}{2}\left[K_c(\partial_x\theta_c)^2 +(\partial_x\phi_c)^2/K_c  \right]\nonumber\\
&-&\frac{\sqrt{2}}{\sqrt{\pi}}\mu \partial_x\phi_{c}- \frac{2u}{\pi^2}\cos{ (\sqrt{8\pi} \phi_{c})}
\end{eqnarray}
and the only quantities that determine asymptotics of the algebraically
decaying correlation functions are spin wave velocity $v_s$ and Luttinger
liquid parameter $K_s$ of the spin mode. Clearly, spin and charge modes are
perfectly separated in (\ref{efHam}). However, away of half filling and for
nonzero magnetic field spin and charge degrees of freedom do not separate any
more. The reason of the admixture between spin and charge degrees of freedom
can be easily traced back to the case of noninteracting electrons. For
non-zero magnetization, $m\neq 0$, the anisotropy of the Fermi velocities
$v_{\uparrow}\neq v_{\downarrow}$ (away of half filling) leads to a coupling
of spin and charge excitations\cite{Penc,Giamarchi,Kollath}. For $u=0$ this coupling is:
\begin{equation}
\sim (v_{\uparrow}-v_{\downarrow})(\partial_x \phi_c \partial_x \phi_s+\partial_x \theta_c \partial_x \theta_s ).
\end{equation}
For all filling (except of half filling) there is a quadratic coupling between
spin and charge modes with an amplitude proportional to the velocity
anisotropy at $u=0$.  The modification of this admixture for finite $u$ is
encoded in the dressed charge matrix.  To take fully into account this
spin-charge coupling one has to work with the renormalized theory with the
initial free value of the coupling amplitude $v_{\uparrow}-v_{\downarrow} \sim
m(n-1)/2$.

\subsection{Non-Perturbative Effective Theory}
The effective field theory of the repulsive Hubbard model in a magnetic field
can be constructed in a non-perturbative way by matching the critical
exponents of the multicomponent Luttinger liquid with those of the Hubbard
model \cite{Penc} in the situation where there are two gapless modes.  Here,
we will carry out this comparison in the limit of the zero doping.  In the
presence of two gapless modes the following asymptotic expression follows for
the primary fields from the hypothesis of multivelocity conformal theories
\cite{IzKR89}:
\begin{equation}
\left<  \phi(\tau,x) \phi(\tau,x)\right>=\frac{\exp(2iD_ck_{F,\uparrow}x)\exp(  2i(D_c+D_s)k_{F\downarrow}x)}{(v_c\tau+ix)^{2\Delta_c}(v_c\tau-ix)^{2\bar \Delta_c} (v_s\tau+ix)^{2\Delta_s}(v_s\tau-ix)^{2\bar \Delta_s} },
\end{equation}
where the anomalous exponents are defined in terms of the dressed charge
matrix as \cite{FrKo90}:
\begin{eqnarray}
2\Delta_c(\bar \Delta_c)&=& \left(Z_{cc}D_c+Z_{sc}D_s+(-)\frac{Z_{ss}\Delta N_c-Z_{cs}\Delta N_s}{2 \det Z}    \right)^2+2N_c(\bar N_c),\\
2\Delta_s(\bar \Delta_s)&=& \left(Z_{cs}D_c+Z_{ss}D_s+(-)\frac{Z_{cc}\Delta N_s-Z_{sc}\Delta N_c}{2 \det Z}    \right)^2+2N_s(\bar N_s).
\end{eqnarray}

At $h=0$ the dressed charge matrix takes the form:
\begin{equation}
Z =\left( \begin{array}{c c}
   Z_{cc}& Z_{cs} \\
 Z_{sc} & Z_{ss}
\end{array} \right)=\left( \begin{array}{c c}
   \sqrt{2K_c } & 0 \\
 \sqrt{K_c/2} & \sqrt{K_s/2}
\end{array} \right),
\end{equation}
with the familiar Luttinger liquid parameters $K_{c/s}$ of charge/spin sectors.
On the other hand we have $\Delta M=\Delta N_c-2\Delta N_{s}$,
$2D_c+D_s=J_c/2k_F$, $D_s=-J_s/2k_F$ \cite{Woyn89}.  Therefore, at zero
magnetic field one finds $\Delta_c (\bar\Delta_c)= f(\Delta N_c, J_c)$ and
$\Delta_s (\bar\Delta_s)= f(\Delta M, J_s)$, i.e.\ the anomalous dimensions of
spin and charge fields depend only on spin and charge quantum numbers
respectively.  This is the essence of spin-charge separation
\cite{Woyn89,FrKo90}.  For finite magnetic field, $h\neq 0$, and below
half--filling, $n\neq 1$, however, this is no longer the case.  Here the
fields that diagonalize the quadratic Hamiltonian (Luttinger liquid fixed
point) are not spin and charge modes, but rather their linear combinations
denoted by $\phi_{\pm}$ \cite{Penc,Cabra}:
\begin{equation}
\label{effective}
H_{eff}=\frac{v_c}{2}\left[(\partial_x\phi_+)^2+(\partial_x\theta_+)^2\right]
+\frac{v_s}{2}\left[(\partial_x\phi_-)^2+(\partial_x\theta_-)^2\right] .
\end{equation}
Again, the quantity that connects those fields with microscopic physical spin
and charge fields is the dressed charge matrix $Z$:
\begin{eqnarray}
\det Z\,\, \phi_+&=&Z_{ss}\frac{\phi_c+\phi_s}{\sqrt{2}}+(Z_{ss}-Z_{cs})\frac{\phi_c-\phi_s}{\sqrt 2}\,,\nonumber\\
\det Z\,\, \phi_-&=&Z_{sc}\frac{\phi_c+\phi_s}{\sqrt
  2}+(Z_{sc}-Z_{cc})\frac{\phi_c-\phi_s}{\sqrt 2}\, ,
\end{eqnarray}
and
\begin{eqnarray}
 \theta_+&=&(Z_{cc}-Z_{sc})\frac{\theta_c+\theta_s}{\sqrt{2}}+Z_{sc}\frac{\theta_c-\theta_s}{\sqrt 2}\, ,\nonumber\\
\theta_-&=&(Z_{ss}-Z_{cs})\frac{\theta_c+\theta_s}{\sqrt 2}-Z_{ss}\frac{\theta_c-\theta_s}{\sqrt 2}.
\end{eqnarray}
Note that decoupling of the spin and charge degrees of freedom in the
effective Hamiltonian (\ref{effective}) requires $\phi_\pm\propto\phi_{c/s}$
and $\theta_\pm\propto\theta_{c/s}$, i.e.\ $Z_{cs}=0$ and $Z_{cc}=2Z_{sc}$
\footnote{The other possibility for decoupling, i.e.\
  $\phi_\pm\propto\phi_{s/c}$ and $\theta_\pm\propto\theta_{s/c}$, amounts to
  a simple relabeling of the gapless modes.  In the Hubbard model where
  charge and spin modes are related to observable quantum numbers the
  corresponding condition  $Z_{cc}=0$ cannot be realized.}.
As we shall see below the first condition holds for general values of $u$ and
$h$ at half-filling while the second is violated for non-zero magnetic field.

The form (\ref{effective}) of the Hamiltonian is very useful for comparison
with the one obtained by ordinary weak- coupling bosonization procedure. In
particular at zero magnetization where spin and charge fields are decoupled we
can directly read off Luttinger liquid parameters.  Indeed, using the explicit
expression of dressed charge matrix for zero magnetization in the limit of
half filling (see matrix (\ref{dressedzero}) below), the effective theory
takes the following form:
\begin{equation}
H_{eff}=\frac{v_c}{2}\left[2(\partial_x\phi_c)^2+\frac{1}{2}(\partial_x\theta_c)^2\right]+\frac{v_s}{2}\left[(\partial_x\phi_s)^2+(\partial_x\theta_s)^2\right]\, ,
\end{equation}
implying that the fixed point values for Luttinger liquid parameters are
$K_c=1/2$, $K_s=1$.  Note however, that this theory is valid only at extremely
low energies, because the charge wave velocity $v_c\to 0$ approaching Mott
phase (see Eq. (\ref{chargevelocity})).

Once the elements of dressed charge matrix are determined, susceptibilities as
well as correlation functions can be calculated from the above effective
theory.  In the following we will study the elements of dressed charge matrix
at the Mott metal-insulator transition point in the presence of magnetic field
by Bethe ansatz method.

\section{Dressed charge matrix at half filling}
Within the Bethe ansatz approach the dressed charge matrix is a quantity which
determines the finite size spectrum and thereby the various susceptibilities
for the model.  For the Hubbard model it is given as:

\begin{equation}
\label{dresscba}
Z = \left( \begin{array}{cc}
   \xi_{cc}(Q) & \xi_{cs}(A)\\
   \xi_{sc}(Q) & \xi_{ss}(A)
	   \end{array}
\right)\,.
\end{equation}
The elements of $Z$ are solutions of a system of coupled integral equations
\cite{Woyn89,FrKo90}:

\begin{equation}
\label{intxi}
\begin{aligned}
  \xi_{cc}(k) &= 1 +  \int_s \mathrm{d}\lambda'\
    \xi_{cs}(\lambda') a_1(\lambda'-\sin k)\,,
\\
  \xi_{cs}(\lambda) &= \int_c \mathrm{d}k' \cos k'\,
    \xi_{cc}(k')a_1(\sin k'-\lambda)
    - \int_s \mathrm{d}\lambda'\ \xi_{cs}(\lambda')
   a_2(\lambda'-\lambda)\,,
\\
  \xi_{sc}(k) &= \int_s \mathrm{d}\lambda'\
    \xi_{ss}(\lambda') a_1(\lambda'-\sin k)\,,
\\
  \xi_{ss}(\lambda) &= 1+ \int_c \mathrm{d}k'\ \cos k'
    \xi_{sc}(k')a_1(\sin k'-\lambda)
   - \int_s \mathrm{d}\lambda'\ \xi_{ss}(\lambda')
   a_2(\lambda'-\lambda)\ .
\end{aligned}
\end{equation}
Here $a_n(x) = \frac{1}{2\pi}\,\frac{2nu}{(nu)^2+x^2}$ and $\int_c =
\int_{-Q}^Q$, $\int_s=\int_{-A}^A$.  In the Bethe ansatz approach the
boundaries $Q$ and $A$ have to be determined as functions of the filling and
magnetization.  At half filling one has $Q\equiv\pi$ while $A$ is fixed by the
condition $\epsilon_1(\pm A)=0$ for the dressed energy of the magnetic
excitations (the chemical potential in the Mott phase is $\mu=0$):
\begin{equation}
\label{inten1}
\begin{aligned}
\epsilon_1(\lambda) &= h  +4u- 4\mathrm{Re}\sqrt{1-(\lambda-iu)^2}
        - \int_s \mathrm{d}\lambda'\ \epsilon_1(\lambda')
   a_2(\lambda'-\lambda)\, .
\end{aligned}
\end{equation}

Using $\int_{-\pi}^\pi \mathrm{d}k\,\cos k\, f(\sin k) = 0$ the integral
equations (\ref{intxi}) simplify:
\begin{equation}
\label{intxi1}
\begin{aligned}
  \xi_{cc}(k) &= 1\ ,\qquad
\xi_{cs}(\lambda) = 0\ ,\\
  \xi_{sc}(k) &= \int_s \mathrm{d}\lambda'\
    \xi_{ss}(\lambda') a_1(\lambda'-\sin k)\ ,\\
  \xi_{ss}(\lambda) &= 1
   - \int_s \mathrm{d}\lambda'\ \xi_{ss}(\lambda')
   a_2(\lambda'-\lambda)\ .
\end{aligned}
\end{equation}
With this the dressed charge matrix $Z$ reads at half filling:
\begin{equation}
\label{dressc}
  Z = \left(\begin{array}{cc} 1 & 0\\ Z_{sc} & Z_{ss}\end{array}\right)\,,
\end{equation}
and the remaining entries are determined as \footnote{Alternatively, one can
  write $Z_{sc} = \int_s \mathrm{d}\lambda\ f(\lambda)$ with $f(\lambda) =
  a_1(\lambda) - \int_s\mathrm{d}\lambda'\ f(\lambda') a_2(\lambda'-\lambda)$,
  see Ref.\,[\onlinecite{Woynarovich91}].  }:
\begin{equation}
\begin{aligned}
 Z_{sc} &= \xi_{sc}(\pi) = \int_s \mathrm{d}\lambda'\
    \xi_{ss}(\lambda') a_1(\lambda')\ ,\\
  Z_{ss} &= \xi_{ss}(A)\ .
\end{aligned}
\end{equation}

For general Hubbard coupling and magnetic field, the integral equations
(\ref{inten1}) and (\ref{intxi1}) for $\epsilon_1$ and $\xi_{ss}$ have to be
solved numerically to obtain the dressed charge matrix (\ref{dressc}) as a
function of the magnetic field or the magnetization.  In the weak and strong
coupling limits, however, one can obtain asymptotic expressions.

For $u\to\infty$ we have the relations $Z_{cc}=1$,
$Z_{sc}=\frac{1}{2}-\frac{m}{n_c}$ for any electron filling $n_c$ and
magnetization $m$ \cite{Frahm}.  Therefore we find for strong coupling:
\begin{equation}
\label{zsc-sc}
  Z_{cc}-2Z_{sc} = 2\frac{m}{n_c} \to 2m\quad\mathrm{for~}n_c\to1\,.
\end{equation}

In the weak coupling regime we analyze the integral equations for half filling
(\ref{inten1}) and (\ref{intxi1}) for sufficiently small
magnetic fields: in this case the boundary $A$ of the remaining integration
$\int_s$ is large.
Rewriting eqs.~(\ref{intxi1}) as:
\begin{equation}
\label{intxi10}
\begin{aligned}
  \xi_{ss}(\lambda) &= \frac{1}{2} + 
        \left\{\int_A^\infty + \int_{-\infty}^{-A}\right\} \mathrm{d}\lambda'\,
        R(\lambda-\lambda') \xi_{ss}(\lambda')\,, \\
  R(\lambda) &= \int_{-\infty}^{\infty} \frac{\mathrm{d}\omega}{2\pi}\,
               \frac{\mathrm{e}^{i\omega\lambda}}{1+\mathrm{e}^{2u|\omega|}}\,,
\end{aligned}
\end{equation}
we can apply Wiener-Hopf techniques to compute $\xi_{ss}(\lambda)$ in the
relevant region $\lambda\ge A\gg u$ where the second integral in the first equation of
(\ref{intxi10}) can be treated as a perturbation.  To leading order one
obtains:  
\begin{equation}
\label{wh00}
\begin{aligned}
  \tilde{y}^+(\omega) &\equiv \int_0^\infty \mathrm{d}x\, \mathrm{e}^{i\omega
                x} 
                \xi_{ss}(A+x) \simeq \frac{1}{\sqrt{2}(\omega+i\epsilon)}
                G^+(\omega)\,,\\
  G^+(\omega) &= \frac{\sqrt{2\pi}}{\Gamma(\frac{1}{2}-i\frac{u\omega}{\pi})}
                \left(-i\frac{u\omega}
                {\pi\mathrm{e}}\right)^{-i\frac{u\omega}{\pi}}  
\end{aligned}
\end{equation}
Starting with (\ref{wh00}) higher order corrections to $\tilde{y}^+(\omega)$
can be computed in an iterative scheme \cite{YaYa66,Woynarovich91,HubbBook}
giving the diagonal element $Z_{ss}$ of the dressed charge matrix:
\begin{equation}
\label{zss-A}
  Z_{ss} = \lim_{\omega\to\infty} (-i\omega) \tilde{y}^+(\omega)
         = \frac{1}{\sqrt{2}} \left( 1 + \frac{u}{2\pi A}+\ldots\right)
\end{equation}
Rewriting the expression for $Z_{sc}$ in a similar way we obtain:
\begin{equation}
\label{zsc-A}
\begin{aligned}
  Z_{sc} &= \frac{1}{2} -\frac{1}{2u}\int_A^\infty\mathrm{d}\lambda
                  \frac{\xi_{ss}(\lambda)}{\cosh(\pi\lambda/2u)}\\
         &= \frac{1}{2} - \frac{1}{u} \tilde{y}^+\left(i\frac{\pi}{2u}\right)
           \mathrm{e}^{-\frac{\pi}{2u}A} 
           + O\left(\mathrm{e}^{-\frac{3\pi}{2u}A}\right)
          \simeq \frac{1}{2} - \sqrt{\frac{2}{\pi\mathrm{e}}}\,
                  \mathrm{e}^{-\frac{\pi}{2u}A}  
\end{aligned}
\end{equation}
Finally, $A$ has to be expressed in terms of the magnetic field or the
magnetization.  Applying the Wiener-Hopf technique to  Eq.~(\ref{inten1}) the
condition $\epsilon_1(A)=0$ gives:
\begin{equation}
  A = \frac{2 u}{\pi} \ln({h_c}/{h})
    - \frac{u}{2\pi} \frac{1}{\ln(h_c/h)}
    + \ldots
\end{equation}
where $h_c = 4\sqrt{2\pi/\mathrm{e}}\, I_1(\pi/2u)$ \cite{HubbBook}.  
In particular, we have $A\to\infty$ for zero magnetic field independent of
$u$.  As a consequence the dressed charge matrix at half filling for any $u$ becomes:
\begin{equation}
\label{dressedzero}
Z = \left( \begin{array}{cc}
   1 & 0\\
   1/2 & \sqrt{2}/2
	   \end{array}
\right)\,.
\end{equation}
which is the strong coupling limit of the expression obtained in the $SU(2)$
invariant case \emph{below} half filling \cite{FrKo90}.
For small fields the leading field-dependence of the dressed charge
matrix is the linear (up to logarithmic corrections): 
\begin{equation}
\label{zsc-B}
Z_{cc}-2Z_{sc} = \frac{1}{2\pi I_1(\pi/2u)} h\,.
\end{equation}
Finally, with the known small field behavior of the magnetization $m=(h/2\pi
v_s)$ and the spinon velocity $v_s = 2 I_1(\pi/2u)/I_0(\pi/2u)$
\cite{HubbBook}, this expression can be written as a function of the
magnetization $m$:
\begin{equation}
\label{zsc-m}
Z_{cc}-2Z_{sc} = \frac{2}{I_0(\pi/2u)} m
         \simeq \frac{2\pi}{\sqrt{u}}\mathrm{e}^{-\pi/2u}\,m
         \quad \mathrm{~for~}u\to0\,.
\end{equation}

\begin{figure}
\begin{center}
  \includegraphics[width=0.8\textwidth]{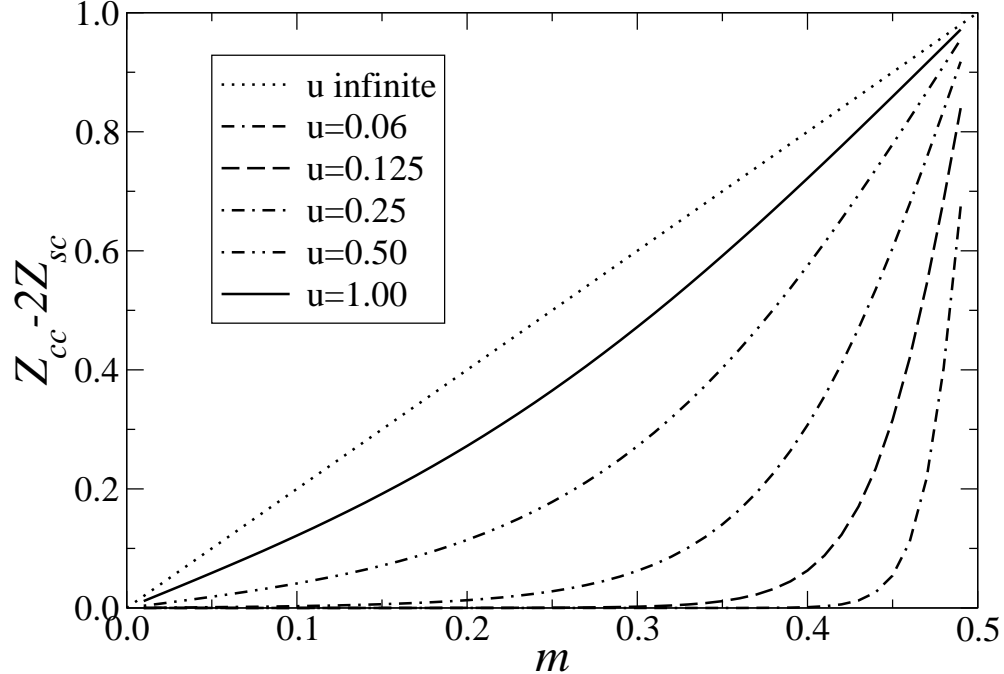}
\end{center}
\caption{\label{fig:zzz}
  The combination $Z_{cc}-2Z_{sc}$ of elements of the dressed charge
  matrix as a function of the magnetization  for chemical potential
  $\mu=\mu_-$ and coupling as indicated.}
\end{figure}

For the crossover between this exponentially suppressed dependence on $u$
\footnote{One can check that the result (\ref{zsc-m}) is consistent with a
  zero magnetization limit of: $({\sin{2 \pi m}}/{\sqrt{u \cos{\pi m}}})
  \mathrm{e}^{-{\pi \cos{\pi m}}/{2u}}$, obtained after particle-hole
  transform from the attractive Hubbard model \cite{Woynarovich91}.}  
to the one for the strong coupling (\ref{zsc-sc}) the Bethe integral equations
(\ref{inten1}) and (\ref{intxi1}) have to be solved numerically.  The result
is shown in Figure~\ref{fig:zzz}.

\section{Susceptibilities and Correlation Functions}
Let us now summarize our results for the elements of the dressed charge matrix
for general magnetization when doping $\delta=1-n_c\to 0$.  As we worked out
explicitly in this limit:

(i)
$Z_{cc}=\sqrt{2K_c}\to 1$: 
this equation simply reflects the fact that holons are extremely dilute in
$n_c \to 1$ limit and therefore they do not interact directly with each other.

(ii)
$Z_{cs}\to 0$:
this holds because this element of dressed charge is a smooth function of
density and magnetization, which is well defined at half filling and any
magnetization. Due to the spin-charge separation it vanishes at half filling.

(iii) $Z_{ss}\to \sqrt{K_s/2}$ depends on magnetization and $u$. In the limit
of infinite $|u|$ it approaches the Luttinger liquid parameter of the
Heisenberg chain at the same magnetization. For weak coupling $|u|\to 0$ we
have:
\begin{equation}
K_s=1+\frac{2u}{\pi v_F}
\end{equation}
where $v_F= 2\sin {\frac{\pi n_c}{2}}$ is the Fermi velocity of the
noninteracting electrons.

(iv) The last element that we worked out was $Z_{sc}$.  In two limits, namely
$u\to 0$ and $u\to \infty$ analytic expressions were obtained.  The most
important result is that, generically, $Z_{cc}-2Z_{sc}\neq 0$ in the limit
$n_c\to 1$, thus the effective theory does not get completely decoupled in
spin and charge parts.  This has important consequences on response functions
and correlation functions.  It is straightforward to study susceptibilities
from our effective theory \cite{Vekua}, and indeed one can see that even at
the Mott metal-insulator transition point the exact expressions for the
Hubbard model
are recovered:
\begin{eqnarray}
\label{susc}
 \chi_c|_{h=const}=\frac{1}{\pi v_c}\to \infty,\,\,\,\,\,\,\,\, \chi_c|_{m=const}=\frac{4}{\pi} \frac{Z^2_{ss}}{v_s(1-2Z_{sc})^2}
\end{eqnarray}
(these and similar expressions for the spin susceptibilities are given in
Ref.\,[\onlinecite{HubbBook}]).  Note that the charge susceptibility remains
finite across the Mott transition when the magnetization is kept constant.

On the other hand when the magnetization is not kept fixed, then the charge
susceptibility diverges across the Mott metal-insulator transition, and we can
estimate how it diverges with the doping. For this we see from
Eq.~(\ref{susc}) that it is sufficient to calculate the charge velocity as a
function of doping.  At half filling in the vicinity of the holon band minimum
there is a relativistic dispersion (Eq.~(7.24) from
Ref.\,[\onlinecite{HubbBook}]): $E(k)=\sqrt{\Delta^2+k^2v^2}$ (corresponding
to expansion of the holon band up to second order in momentum) where $v$ is
the holon velocity at half filling and $\Delta$ is a single particle gap.
Therefore, below half filling the charge velocity can be obtained as\cite{Giamarchi}:
%
\begin{equation}
\label{chargevelocity}
v_c=\left.\frac{\partial E}{\partial k}\right|_{k=\pi \delta}
   =\frac{v^2\delta \pi}{\sqrt{\Delta^2+v^2\pi^2\delta^2}}\,.
\end{equation}
In particular for $u\to \infty$ we have $v=2\sqrt{u}$ and it follows $
v_c=2\pi\delta$.

Once the elements of dressed charge matrix are determined one can easily
obtain correlation functions of physical operators
\cite{FrKo90,Frahm,Penc,Cabra}.  We will discuss below only those correlators,
which decay algebraically on both sides of phase transition.  For the Hubbard
model those correlators involve only spin operators \cite{EsFr99}. The most
slowly decaying one in the Mott insulator phase is the in plane spin-spin
correlation function. In the metallic phase, i.e.\ the phase with two gapless
modes, the leading correlator is, depending on the strength of $u$, either
single field (for small values of $u$) or again in-plain spin correlator (for
large values of $u$).  The equal time correlation function of in plane spin
operator reads:
\begin{equation}
 G_{\sigma;\sigma}^{\perp}(\mu>\mu_{cr})=\frac{e^{2ik_Fx}}{x^{(Z_{cc}^2+Z_{cs}^2)(1/2+1/2(\det Z)^2) }}
\end{equation}
while in the insulating phase it was given by:
\begin{equation}
 G_{\sigma;\sigma}^{\perp}(\mu<\mu_{cr})=\frac{e^{2ik_Fx}}{x^{Z_{cc}^2/2(\det
 Z)^2 }}=\frac{e^{2ik_Fx}}{x^{\frac{1}{K_s}} }\,. 
\end{equation}
Here, we can read off the jump in the critical exponent across the Mott
metal-insulator transition as:
\begin{equation}
 \nu=\frac{Z_{cc}^2}{2}+Z_{cs}^2(\frac{1}{2}+\frac{1}{2(\det Z)^2}) =
 \frac{1}{2}
\end{equation}
Note, that although there is no spin-charge separation the same universal jump
of $1/2$ in critical exponent as in the zero-field case is recovered.

\section{Conclusions}
The presence of a magnetic field breaks time reversal invariance and thereby
prevents the separation of spin and charge degrees of freedom in the 1D
Hubbard model away from half-filling.  In this paper we have studied the
effect of the resulting admixture of the gapless modes on the nature of the
Mott metal-insulator transition.
We have addressed this problem by studying the long wave-length properties of
1D Hubbard model across the transition in an external magnetic field.  Using
Bethe ansatz techniques we calculated numerically elements of dressed charge
matrix for generic cases. In addition exact analytic expressions were obtained
in the limiting cases for the same quantity.  We also constructed a
non-perturbative effective field theory where the drastic effects produced by
the time reversal symmetry breaking across the Mott phase transition become
manifest:
while the susceptibility related to the charge mode which becomes gapped at
the transition diverges at half filling for fixed magnetic field (just as in
the time reversal invariant case and as known for the free fermion picture
\cite{JN} for the commensurate-incommensurate transition) it remains finite
due to the absence of spin charge separation when the magnetization is kept
constant.  The reason for the behavior in the latter case  is that there is
always an admixture to the effective field of the magnetic mode which remains
gapless across the transition.

\section{Acknowledgments}
TV acknowledges discussions with Gora Shlyapnikov who motivated his interest
in the problem. The work was started while TV's visit to the University of
Hanover supported by the Deutsche Forschungsgemeinschaft. TV also
acknowledges GNSF grant No. N\,$06_-81_-4_-100$. LPTMS is a mixed research
unit No.\,8626 of CNRS and Universit\'e Paris Sud.

\end{document}